# Dissipation of quantum information near the event horizon of Schwarzschild black holes

D. Ahn* and S. W. Hwang†

*Institute of Quantum Information Processing and Systems, University of Seoul, Seoul 130-743, Republic of Korea*

† *Department of Electronics Engineering, Korea University, Seoul 136-075, Republic of Korea*

*Abstract:* We found that the Wigner rotation of a particle with spin 1/2, which is unitary in Minkowski spacetime, becomes non-unitary as the particle is falling onto the black hole. This implies that, from the quantum information processing point of view, any quantum information encoded in spins will be dissipated near the black hole. The gravitational field around the black hole thus acts like a dissipative quantum channel, especially the bit flip channel.

* E-mail: dahn@uos.ac.kr

The information paradox for black holes [1-7] has been one of the major problems in theoretical physics for the last thirty years. The problem is that black holes appear to eat the quantum information as well as the matter, yet the most fundamental laws of physics demand that this information should be preserved as the universe evolves. In the case of scalar particles, it is well known [8-10] that quantum information is dissipated by Hawking-Unruh effect [1-3]. On the other hand, for the fermion excited state (or a particle state) the incoming and outgoing Hawking radiation are decoupled [11,12]. This immediately suggests that the behaviour of a spin under the influence of the strong gravity near the event horizon of a black would play a key role in the fate of the quantum information.

Here, we considered spin 1/2 particles falling onto the Schwarzschild black holes and studied the change of quantum correlation encoded in spins from the quantum information processing point of view. Describing the motion of the particle with quantum mechanical spin near the black hole is non trivial because the Poincare group does not act naturally in the curved spacetime. In Minkowski spacetime, the change of spin polarization, as a result of the particle's motion, is described by the Wigner rotation [13-19]. We first generalized the Wigner rotation to the case of spin in local inertial frames for an infinitesimal displacement in curved spacetime.

The Wigner rotation is essentially a quantum operation. According to the fundamental theorem of quantum information processing, the initial quantum correlation or quantum information encoded in spins is preserved only if the quantum operation is unitary [15]. Then the question of whether the quantum information stored in the spin of the particle is preserved as the particle falls onto the event horizon of the black hole, is reduced to the problem of determining if the resulting Wigner rotation is unitary. When a particle is under the influence of a strong gravity, it is found that the Wigner rotation is non-unitary due to the spin-orbit interaction. This implies that, from the quantum information processing point of view, any quantum information stored in the particle would not be preserved near the event horizon of the black hole. The gravitational field near the black hole thus acts like a dissipative quantum channel [21-23], more specifically a bit flip channel. The information is dissipated into the gravitational field or more specifically the metric of the curved spacetime.

*-Wigner rotation in curved spacetime:* We proceed to consider the motion of spin 1/2 particles falling onto the event horizon of the black hole. In order to determine the effects of gravitation on general physical system, we replace all Lorentz tensors, which describe the given special-relativistic equations in Minkowski spacetime, with objects that behave like tensors under general coordinate transformations [24,25]. Also, we replace all derivatives with covariant derivatives and replace Minkowski metric tensor $\eta_{ab}$ with the metric tensor $g_{\mu\nu}$. The equations are then generally covariant. This method, however, works only for objects that behave like tensors under Lorentz transformation and not for the spinor fields. Especially, the description of spin requires the introduction, at each point, of an independent Lorentz coordinate frame, combined with the demand of invariance under local Lorentz transformations [26]. The relation between the local and the general coordinate system is conveyed by a family of vector fields called tetrads $e_a^\mu(x)$, $a = 0, 1, 2, 3$, which respond to general coordinate transformations and local Lorentz transformations as [26]

$$\bar{e}_a^\mu(\bar{x}) = \frac{\partial \bar{x}^\mu}{\partial x^\nu} e_a^\nu(x) \quad \text{and} \quad \bar{e}_a^\mu(x) = \Lambda_a^{\ b} e_b^\mu(x),$$

respectively. Moreover, we have $g_{\mu\nu} = e_\mu^a e_\nu^b \eta_{ab}$. Since the tetrad $\bar{e}_a^\mu(x)$ is a vector field representing a tangent plane to general coordinates, two such vectors at different points can not be compared naively with each other. For example, in order to subtract $\bar{e}_a^\mu(x + \delta x)$ from $\bar{e}_a^\mu(x)$, we have to transport $\bar{e}_a^\mu(x)$ from $x$ to $x + \delta x$ without change and compare the difference [27]. This transport of vector is known as parallel transport. For an infinitesimal coordinate transformation, we get

$$\bar{e}_a^\mu(x + \delta x) - \bar{e}_a^\mu(x) \rightarrow \delta x^\lambda \frac{\partial}{\partial x^\lambda} e_a^\mu(x) + \delta x^\lambda e_a^\nu(x) \Gamma_{\nu\lambda}^\mu(x) = \delta x^\lambda \nabla_\lambda e_a^\mu(x), \quad (1)$$

where $\Gamma_{\nu\lambda}^\mu$ is the affine connection and $\nabla_\lambda$ is the covariant derivative [24-27]. We also have

$$\begin{aligned}
\delta x^\lambda \nabla_\lambda e_a^\mu &= \delta x^\lambda e_b^\mu e_\kappa^b \nabla_\lambda e_a^\kappa \\
&= \delta x^\nu e_\nu^c \Gamma_{ca}^b e_b^\mu, \\
&= \delta \omega^b{}_a e_b^\mu
\end{aligned} \quad (2)$$

with $\Gamma_{ca}^b = e_\kappa^b e_c^\lambda \nabla_\lambda e_a^\kappa$ and $\delta \omega^b{}_a = \delta x^\nu e_\nu^c \Gamma_{ca}^b = -\delta \omega_a{}^b$. Here, we have used the following orthogonal properties of tetrad $e_a^\mu(x)$ and its inverse $e_\mu^a(x)$: $e_a^\mu(x) e_\nu^a(x) = \delta_\nu^\mu$, $e_a^\kappa(x) e_\kappa^b(x) = \delta_a^b$. Since any vector V is independent of the basis, it follows that $V^\mu = e_a^\mu V^a$ and $V^a = e_\mu^a V^\mu$. From equations (1) and (2), we also note that an infinitesimal displacement in curved spacetime can be viewed as the Lorentz transformation in the local inertial frame [26]. It is straightforward to show that $\delta\omega_{ba} = -\delta\omega_{ab}$, so the infinitesimal local Lorentz transformation is give by

$$\delta e_a^{\mu}(x) = \delta\omega_a{}^b(x)\, e_b^{\mu}(x), \quad \Lambda_a{}^b(x) = \delta_a^b + \delta\omega_a{}^b(x) \quad . \tag{3}$$

Throughout the paper, we are using the Latin letters for four local inertial coordinates and the Greek letters for general coordinates.

The Hilbert space vector for a spin 1/2 particle is then defined on the local inertial frame spanned by the tetrads. On the tangent plane defined by the basis $\{e_a^{\mu}(x)\}$, an infinitesimal Lorentz transformation $\Lambda$ induced by a coordinate displacement transforms vectors in the Hilbert space as $\Psi|_x \to U(\Lambda)\Psi|_{x'}$ [17,18]. The quantum state has the following transformation rule [17,18]:

$$|\vec{p},\sigma\rangle \to \sum_{\bar{\sigma}} D^{(j)}_{\bar{\sigma}\sigma}(W(\Lambda,p^a))|\vec{p}_{\Lambda},\bar{\sigma}\rangle. \tag{4}$$

Here, $W(\Lambda,p)$ is Wigner's little group element, $D^{(j)}(W)$ the representation of $W$ for spin $j$, $p^a = (\vec{p},p^0)$, $(\Lambda p)^a = (\vec{p}_{\Lambda},(\Lambda p)^0)$ with $a = 0, 1, 2, 3$, and $L(p)$ is the Lorentz transformation $p^a = L^a{}_b k^b$, and $k^b = (m, 0, 0, 0)$ is the four-momentum taken in the particle's rest frame. In general, unlike the Minkowski spacetime, defining the four momenta in curved spacetime is nontrivial. Nonetheless, one can still define the four momenta from the positive frequency components of the quantum field namely those solutions whose Fourier transform on $J_+$ of the Kruskal extension of Schwarzschild spacetime (Fig. 1) contains only positive frequencies [28]. One also needs to be careful in interpreting the equation (4). The infinitesimal displacement causes the spin of the particle transported to one local inertial frame to another. As a result, the Wigner representation may not necessarily be unitary. Throughout the paper we use units $c = G = 1$. The signature of the metric is defined as $(-, +, +, +)$.

*-Wigner rotation in the Kruskal extension of Schwarzschild spacetime:* We now consider the case when a spin 1/2 particle is falling onto the event horizon of a Schwarzschild black hole. The stationary state for such a black hole is represented by the metric

$$ds^2 = -\left(1 - \frac{2M}{r}\right)dt^2 + \frac{dr^2}{1 - \frac{2M}{r}} + r^2\left(d\theta^2 + \sin^2\theta\, d\varphi^2\right),$$

where $M$ is the mass of the black hole. At $r = 2M$, the Schwarzschild spacetime has an event horizon. The general coordinate is $x^\mu = (t, r, \theta, \varphi)$ with the metric tensor given by

$$g_{tt} = -\left(1 - \frac{2M}{r}\right), \quad g_{rr} = \frac{1}{1 - \frac{2M}{r}}, \quad g_{\theta\theta} = r^2, \quad g_{\varphi\varphi} = r^2 \sin^2\theta,$$

$$g_{\mu\nu} = e^a_\mu e^b_\nu \eta_{ab},$$

where $\eta_{ab}$ is the Minkowski metric tensor with the following components:

$$\eta_{00} = -1, \quad \eta_{11} = \eta_{22} = \eta_{33} = 1.$$

At $r = 2M$, the Schwarzschild spacetime has an event horizon. For Schwarzschild spacetime, one can choose the following tetrads and their inverses [19]:

$$e^0_t = \sqrt{1 - \frac{2M}{r}}, \quad e^1_r = \frac{1}{\sqrt{1 - \frac{2M}{r}}}, \quad e^2_\theta = r, \quad e^3_\varphi = r\sin\theta, \tag{5a}$$

and

$$e^t_0 = \frac{1}{\sqrt{1 - \frac{2M}{r}}}, \quad e^r_1 = \sqrt{1 - \frac{2M}{r}}, \quad e^\theta_2 = \frac{1}{r}, \quad e^\varphi_3 = \frac{1}{r\sin\theta}. \tag{5b}$$

However, in this coordinate the Killing vector is not be defined at the even horizon and as a result, the positive frequency components of the field are not well defined near the horizon. These difficulties can be avoided if we use the Kruskal extension of the Schwarzschild spacetime [25]. In the Kruskal coordinate coordinates, we make the following transformations:

$$X = \left(\frac{r}{2M} - 1\right)^{1/2} \exp\left(\frac{r}{4M}\right) \cosh\left(\frac{t}{4M}\right), \tag{6a}$$

and

$$T = \left(\frac{r}{2M} - 1\right)^{1/2} \exp\left(\frac{r}{4M}\right) \sinh\left(\frac{t}{4M}\right). \tag{6b}$$

The Schwarzschild metric becomes

$$ds^2 = \frac{32M^3}{r} \exp\left(-\frac{r}{2M}\right)\left(-dT^2 + dX^2\right) + r^2\left(d\theta^2 + \sin^2\theta d\varphi^2\right), \tag{7}$$

with $X^2 - T^2 > -1$. Tetrads in the Kruskal extension are given by

$$e^0_T = \left(\frac{32M^3}{r}\right)^{1/2} \exp\left(-\frac{r}{4M}\right), \quad e^1_X = \left(\frac{32M^3}{r}\right)^{1/2} \exp\left(-\frac{r}{4M}\right), \quad e^2_\theta = r, \quad e^3_\varphi = r\sin\theta,$$

and

$$e_o{}^T = \left(\frac{r}{32M^3}\right)^{1/2} \exp\left(\frac{r}{4M}\right), e_1{}^X = \left(\frac{r}{32M^3}\right)^{1/2} \exp\left(\frac{r}{4M}\right), e_2{}^\theta = \frac{1}{r}, e_3{}^\varphi = \frac{1}{r\sin\theta}. \quad (8)$$

The rest of the components is zero. The Killing vector is then given by $\frac{1}{4M}\left(X\frac{\partial}{\partial T} + T\frac{\partial}{\partial X}\right)$, which time like for $|X|>|T|$ and spacelike for $|X|<|T|$ [29].

In order to find the infinitesimal Lorentz transformation on the local inertial frame, one needs to calculate the connection-one form $\delta\omega^a{}_b(x)$ [26,27]. From equations (1), (2) and (8), we obtain:

$$\delta\omega^0{}_1 = -\frac{1}{X}\left(\frac{2M}{r}-1\right)^2 dT, \quad \delta\omega^2{}_0 = \left(\frac{2r}{2M}\right)^{1/2}\frac{\left(1-\frac{2M}{r}\right)}{T}\exp\left(\frac{r}{4M}\right)d\theta,$$

$$\delta\omega^3{}_0 = \left(\frac{r}{2M}\right)^{1/2}\frac{\left(1-\frac{2M}{r}\right)}{T}\exp\left(\frac{r}{4M}\right)\sin\theta d\varphi, \quad \delta\omega^3{}_2 = \cos\theta d\varphi,$$

$$\delta\omega^2{}_1 = \left(\frac{r}{2M}\right)^{1/2}\frac{\left(1-\frac{2M}{r}\right)}{X}\exp\left(\frac{r}{4M}\right)d\theta, \quad \delta\omega^3{}_1 = \left(\frac{r}{2M}\right)^{1/2}\frac{\left(1-\frac{2M}{r}\right)}{X}\exp\left(\frac{r}{4M}\right)\sin\theta d\varphi.$$
$$(9)$$

Since both tetrads and Killing vector are well defined even at the event horizon the infinitesimal Lorentz transformation, induced by the coordinate transformation should preserve the timelike Killing vector and the homogeneous transformation given by equation (4) is well defined.

We also define the four-velocity $U^\mu$ of a particle to be the unit tangent, as measured by $g_{\mu\nu}$, to its world line and we consider the following simple case:

$$U^t = \frac{\cosh\alpha}{\sqrt{1-\frac{2M}{r}}}, \quad U^r = \sqrt{1-\frac{2M}{r}}, \quad U^\varphi = U^\theta = 0,$$

which corresponds to the particle falling straight into the black hole. In this simple model we ignored the spiral motion of the particle.

Then the four-velocity in the local inertial frame becomes $U^0 = \cosh(\alpha + t/4M), U^1 = \sinh(\alpha + t/4M), U^2 = U^3 = 0$.

The Wigner representation $D^{1/2}(W(\Lambda,p)(x))$ for the local inertial frame is derived using the same approach as the case of special relativity [17,18]:

$$D^{1/2}(W(\Lambda,p)(x))$$

$$= \begin{pmatrix} 1 - \dfrac{K^2 \coth\beta}{(1-K^2)X} bdT & -\dfrac{(1+2K\coth\beta-K^2)bdT}{2(1-K^2)X} \\ +i\dfrac{(2KX-T-K^2[3T+2X\coth\beta])ad\theta}{2(1-K^2)XT} & -\dfrac{i}{2}\left(d\varphi - \dfrac{([1-2K\coth\beta+K^2]-4KT)}{2(1-K^2)XT}ad\theta\right) \\ -\dfrac{(1+2K\coth\beta-K^2)bdT}{2(1-K^2)X} & 1 - \dfrac{K^2\coth\beta}{(1-K^2)X}bdT \\ -\dfrac{i}{2}\left(d\varphi + \dfrac{([1-2K\coth\beta+K^2]-4KT)}{2(1-K^2)XT}ad\theta\right) & -i\dfrac{(2KX-T-K^2[3T+2X\coth\beta])ad\theta}{2(1-K^2)XT} \end{pmatrix}.$$

(10)

Here $a = \left(\dfrac{r}{2M}\right)^{1/2}\left(1-\dfrac{2M}{r}\right)\exp\left(\dfrac{r}{4M}\right)$, $b = \left(\dfrac{2M}{r}-1\right)^2$, $\beta = \alpha + t/4M$, and

$$K = \left(\dfrac{p^0 - m}{p^0 + m}\right)^{1/2}, \quad 0 < K < 1.$$

The four-momentum is given by $p^a = mU^a = (p^0, \vec{p})$. Here m is the mass of the particle and $\cosh\alpha = \sqrt{\vec{p}^2 + m^2}/m$. These results can be extended to the finite displacement by successive applications of the infinitesimal Wigner rotation. It is now obvious that the Wigner rotation given by equation (10) is not unitary since the general unitary matrix $U(A,B)$ should have the following form [27]:

$$U(A,B) = \begin{pmatrix} A & B \\ -B* & A* \end{pmatrix}, \quad (11)$$

where $A$ and $B$ are complex quantities and * is the complex conjugate. Wigner rotation becomes unitary when $K \to 0$, and $X \to \infty$, which is nothing but the flat spacetime condition. In this case, the Wigner rotation is given by

$$D^{1/2}(W(\Lambda,p)) \to \begin{pmatrix} 1 & \dfrac{-id\varphi}{2} \\ -\dfrac{id\varphi}{2} & 1 \end{pmatrix}. \quad (12)$$

In this case, the Wigner rotation is unitary up to the first order and the quantum correlation is preserved throughout the infinitesimal displacement. In the following, we ignore the effect of the angular coordinate by assuming $d\theta = 0$ and $d\varphi = 0$. For a finite $T$, we iterate the infinitesimal transformation and obtain

$$D^{1/2}(W(\Lambda,p)) = \begin{pmatrix} T_c \exp\left(-\int \dfrac{K^2 \coth\beta}{(1-K^2)X} bdT\right) & T_c \exp\left(-\int \dfrac{1+K \coth\beta - K^2}{2(1-K^2)X} bdT\right) - 1 \\ T_c \exp\left(-\int \dfrac{1+K \coth\beta - K^2}{2(1-K^2)X} bdT\right) - 1 & T_c \exp\left(-\int \dfrac{K^2 \coth\beta}{(1-K^2)X} bdT\right) \end{pmatrix},$$

(13)

where $T_c$ is the time ordering operator with respect to $T$.

***-Dissipation of quantum information in gravitational field:*** Now, let's consider the transformation of spin in the gravitational field. Let us assume that the particle's spin information be described by the density operator $\rho$. Then the result of the Wigner rotation on the particle's spin is given by

$$\rho' = D^{1/2}(W(\Lambda,p)) \rho \left(D^{1/2}(W(\Lambda,p))\right)^\dagger,$$
$$= (pI - (1-q)\sigma_1) \rho (pI - (1-q)\sigma_1)$$

(14)

where $I$ is the unit matrix, $\sigma_1 = \begin{pmatrix} 0 & 1 \\ 1 & 0 \end{pmatrix}$, $p = T_c \exp\left(-\int \dfrac{K^2 \coth\beta}{(1-K^2)X} bdT\right)$, and $q = T_c \exp\left(-\int \dfrac{1+K \coth\beta - K^2}{2(1-K^2)X} bdT\right)$. If we assume that the particle is spin is initially aligned to the 3-direction, i.e., $\rho = \begin{pmatrix} 1 & 0 \\ 0 & 0 \end{pmatrix}$, the resulting density operator becomes

$$\rho' = \begin{pmatrix} p^2 & -p(1-q) \\ -p(1-q) & (1-q)^2 \end{pmatrix}.$$

(15)

In quantum information theory, the dissipation of quantum information in a noisy channel is described by von Neumann entropy $S(\rho) = -Tr(\rho \log \rho)$ which indicates the uncertainty associated with a given quantum state [20]. In the present case $S(\rho) = 0$, denoting the initial state is a pure state. On the other hand, when a particle is under the gravity, von Neumann entropy is $S(\rho') = -(p^2 + (1-q)^2)\log[p^2 + (1-q)^2]$, which is non-zero in general. It is interesting to note that equation (14) is similar to the case of bit flip channel [20].

This behavior of spin under the gravity may have something to do with the spin-orbit coupling due to the gravitation. For classical spinning particle in curved space, the momentum and the spin are related by the set of equations [30]:

$$\frac{dx^\mu}{d\tau} = v^\mu, \quad \frac{Dp^\mu}{D\tau} = -\frac{1}{2} R^\mu{}_{\nu\rho\sigma} S^{\rho\sigma}, \quad \frac{DS^{\mu\nu}}{D\tau} = p^\mu v^\nu - p^\nu v^\mu, \quad p^\mu S_{\mu\nu} = 0.$$

In this case, the spin-orbit interaction is repulsive if the spin is parallel to the orbital angular momentum and it would balance with the gravity and if the direction of the spin is opposite, the effect is reverse. Such a spin-orbit interaction is induced through the gravitational interaction. For quantum spin, we do not have such a set of equations yet. Nevertheless, we would like to note that from equation (9) one can see that the infinitesimal displacement of a particle gives the boost along the 1-axis which is inward normal to the event horizon. Thus part of the momentum change may be balanced by the change of spin if we require $p_a S^a = 0$ in the local inertial frame as in the case of classical spinning particle in curved spacetime.

In summary, we found that the Wigner rotation of a particle with spin 1/2, which is unitary for Minkowski spacetime, becomes non-unitary near the black hole. This implies that, from the quantum information processing point of view, any quantum information encoded in the spins will be dissipated as the particle approaches to the event horizon. The gravitational field around the black hole acts like a bit flip quantum.

***Acknowledgements:*** D. A. thanks Y. H. Moon for discussions. This work was supported by the Korean Ministry of Science and Technology through the Creative Research Initiatives Program under Contract No. M1-0116-00-0008

**Figure legend**

**Figure 1** The Kruskal exptention of the Schwarzschild spacetime [25]. In region *I*, null asymptotes $H_+$ and $H_-$ act as futue and past event horizons, respectively. The boundary lines labelled $J^+$ and $J^-$ are fute and past null infinities, respectively, and $i^o$ is the spacelike infinity.

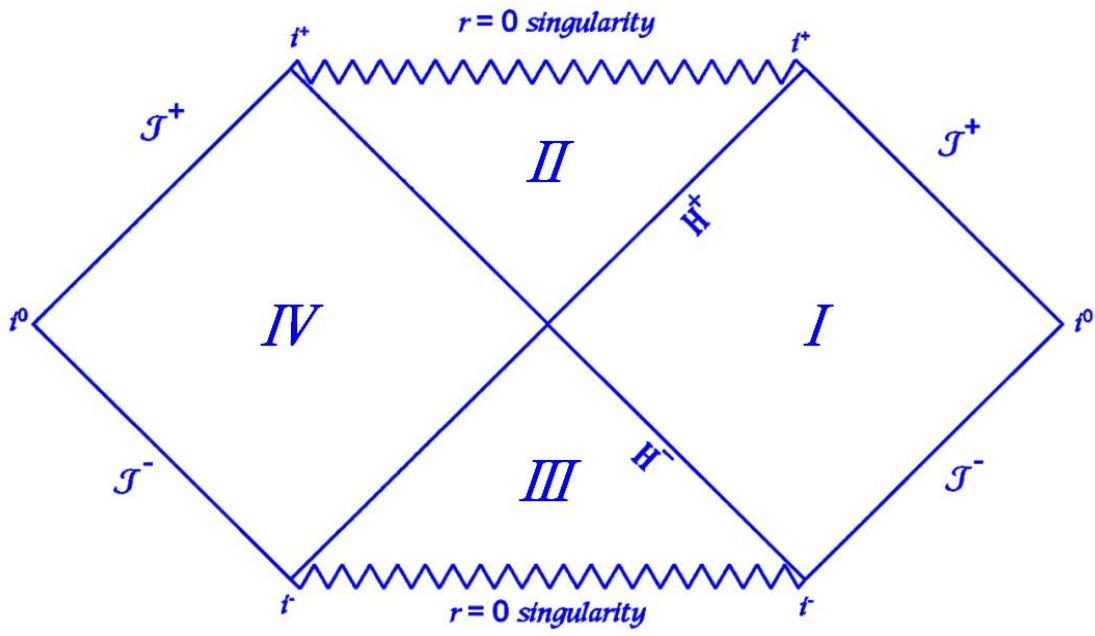

Fig. 1